\documentclass[a4paper,conference]{IEEEtran}

\usepackage[utf8]{inputenc}
\usepackage[T1]{fontenc}
\usepackage{url}
\usepackage{ifthen}
\usepackage{cite}
\usepackage[cmex10]{amsmath} 
\usepackage{inputenc, graphicx, verbatim,color}
\usepackage{float}
\usepackage{tikz}
\usepackage{url}
\usepackage{xcolor}
\usepackage{bbm}
\usepackage[justification=centering]{caption}

\usepackage{amsfonts,amssymb,amsbsy,amsthm}

\newtheorem{theorem}{Theorem}
\newtheorem{corollary}{Corollary}
\newtheorem{lemma}{Lemma}
\newtheorem{proposition}{Proposition}
\newtheorem{definition}{Definition}

\newtheorem{example}{Example}

\newcommand{\F}{\mathbb{F}}
\newcommand{\Z}{\mathbb{Z}}

\newcommand{\LL}{\mathcal{L}}
\newcommand{\ZZ}{\mathcal{Z}}

\newcommand{\Aa}{\mathcal{A}}
\newcommand{\cyc}{\mathrm{cyc}}
\newcommand{\cl}{\mathrm{cl}}
\newcommand{\ie}{\emph{i.e.}}

\newcommand{\st}{\text{ such that }}
\newcommand{\nd}{\text{and}}
\newcommand{\defeq}{:=}

\newcommand\SmallMatrix[1]{{%
  \small\arraycolsep=0.6\arraycolsep\ensuremath{\begin{pmatrix}#1\end{pmatrix}}}}

\begin{document}
\title{Bounds on Binary  Locally Repairable Codes Tolerating Multiple Erasures}


 \author{%
   \IEEEauthorblockN{Matthias Grezet\IEEEauthorrefmark{1}, Ragnar Freij-Hollanti\IEEEauthorrefmark{2}, Thomas Westerb\"{a}ck\IEEEauthorrefmark{1}, Oktay Olmez\IEEEauthorrefmark{3} and Camilla Hollanti\IEEEauthorrefmark{1}}
   \IEEEauthorblockA{\IEEEauthorrefmark{1}%
              Department of Mathematics and Systems Analysis, 
              Aalto University, 
              Finland,
              Email: firstname.lastname@aalto.fi}
   \IEEEauthorblockA{\IEEEauthorrefmark{2}%
              Department of Electrical and Computer Engineering, 
              Technical University of Munich, 
              Germany,
              Email: ragnar.freij@tum.de}
   \IEEEauthorblockA{\IEEEauthorrefmark{3}%
              Department of Mathematics, 
              Ankara University, 
              Turkey,
              Email: oolmez@ankara.edu.tr}
              }

\maketitle

\begin{abstract}
Recently, locally repairable codes has gained significant interest for their potential applications in distributed storage systems. However, most constructions in existence are over fields with size that grows with the number of servers, which makes the systems computationally expensive and difficult to maintain. Here, we study linear locally repairable codes over the binary field, tolerating multiple local erasures. We derive bounds on the minimum distance on such codes, and give examples of LRCs achieving these bounds. Our main technical tools come from matroid theory, and as a byproduct of our proofs, we show that the lattice of cyclic flats of a simple binary matroid is atomic. 
\end{abstract}

\section{Introduction}

In modern distributed  storage systems (DSSs) failures happen frequently, whence decreasing the number of connections required for node repair is crucial. Removing even one connection locally can easily imply huge gains in the overall system functionality, thanks to shortened queues and improved data availability. 
Consequently, locally repairable codes (LRCs) have gained a lot of interest in the past few years \cite{gopalan12, tamo13, tamo14, wang14b}. Namely, LRCs allow to repair a small number of failures locally, \emph{i.e.}, by only contacting few close-by nodes and hence avoiding congesting the system. Related Singleton-type bounds have been derived for various cases, see \cite{tamo16, pollanen16, wang15}. The first bound on the minimum distance for fixed field size was obtained by Cadambe and Mazumdar in~\cite{cadambe15}. Recently, this bound was improved and generalized, via the observation that any log-convex bound on the ``local rank'' of a code can be blown up to obtain a bound on the global rank~\cite{agarwal17}. Interestingly, the bounds in~\cite{agarwal17} do not depend on the linearity of the code. However, all the bounds in~\cite{cadambe15, agarwal17} are implicit, except for special classes of codes. 

In this paper, we consider binary codes motivated by the fact that the computational complexity when retrieving a file or repairing a  node grows with the field size. We derive new, improved Singleton-type bounds for this special case alongside with sporadic examples, in particular when the local repair sets can tolerate multiple failures. In contrast to the bounds in~\cite{cadambe15, agarwal17}, our bounds are explicit, and do not depend on any prior bounds on binary codes of shorter length. 

As our main contribution, in Theorem~\ref{thm:new_dsb_noalpha}, we obtain a closed-form bound on the minimum distance $d$ of a binary $(n,k)$-code of length $n$ and dimension $k$ and with all-symbol $(r,\delta)$-locality, where the local distance $\delta>2$. Such bounds were previously only known when $\delta=2$. The bound is in terms of the rank $\ell$ of the repair sets, but can easily be transformed to bounds in terms of the size $r+\delta-1$. Interestingly, while the two parameters $r$ and $\ell$ can be assumed to agree when $\delta=2$, as well as when the field size is unbounded, this is no longer the case over the binary field with $\delta>2$. While both parameters are of independent interest in applications, we have chosen to focus on the number of nodes $\ell$ that need to be contacted for local repair, rather than on the size of the local clouds\footnote{Note that here, while $r$ still essentially bounds the size of the local sets, we never need to contact that many nodes, since $\ell\leq r-1$ as shown in Prop.~\ref{prop:bounds_repairsets}. From another point of view, $r$ can be seen as a parameter describing ``how far'' one needs to go in order to repair.}.

In addition, in Section~\ref{sec:lattice} we prove that every element of a non-degenerate binary locally repairable code without replication is contained in an atomic cyclic flat, and hence that the lattice of cyclic flats is atomic. From a practical point of view, this implies a hierarchy of failure tolerance, as explained in the end of Section~\ref{sec:lattice}. In particular, whenever a symbol $e$ fails, we can start by downloading nodes in an atomic cyclic flat in order to repair $e$. If it turns out that some other nodes in this local set have failed as well, we can repair them while still keeping the part that we already downloaded, and simply contact some more nodes in the corresponding repair set to repair all the failed symbols. Thus, we do not have to restart from the beginning if we find out during the repair process that a small amount of other nodes have failed as well.

Several constructions are known for optimal LRCs over the binary field, for specified ranges of parameters, and almost exclusively in the case $\delta=2$. The first such construction, for codes with exponentially low rate and locality $r=2,3$, was obtained by deleting carefully chosen columns from the simplex code\cite{silberstein15}. These constructions are also optimal when taking the availability $t$, \emph{i.e.}, the number of disjoint sets that can recover a given symbol, into consideration. A slightly more flexible family of codes, allowing for higher rate, was given in~\cite{huang15, huang16}, where also a slight improvement over the Cadambe-Mazumdar bound was given for linear codes. In the realm of multiple erasures, \ie,  when $\delta>2$, rate-optimal codes were studied in~\cite{balaji16}. There, rate-optimal codes were characterized when $\delta=3$, and analogous constructions without optimality proof were given for $\delta>3$. However, to the best of our knowledge, no previous work has studied bounds on the global minimum distance in the regime $\delta>2$.

\section{Preliminaries of LRCs and Matroids}

As is common practice, we say that $C$ is an $(n,k,d)$-code if it has length $n$, dimension $k$, and minimum Hamming distance $d$. A linear $(n,k,d)$-code $C$ over a finite field $\F$ is a \emph{non-degenerate storage code} if $d \geq 2$ and there is no zero column in a generator matrix of $C$. To study LRCs in more detail, we consider punctured codes $C|Y$, where $Y\subseteq [n]$ is a set of coordinates of the code $C$. For a fixed code $C$, we denote by $d_Y$ the minimum Hamming distance of the punctured code $C|Y$.

\begin{definition}
Let $C$ be an $(n,k,d)$-code over $\F^{n}$. A symbol $x \in [n]$ has \emph{locality} $(r,\delta)$ if there exists a subset $R$ of $[n]$, called \emph{repair set} of $x$, such that $x \in R$, $|R|\leq r+\delta-1$, and $d_R\geq \delta$.
\end{definition}

\begin{definition}
A \emph{linear $(n,k,d,r,\delta)$-LRC} over a finite field $\F^{n}$ is a non-degenerate linear $(n,k,d)$-code $C$ over $\F^n$ such that every coordinate $x \in [n]$ has locality $(r,\delta)$. In the literature, this is specifically called \emph{all-Symbol locality}.  
\end{definition}



The parameters $(n,k,d,r,\delta)$ can immediately be defined and studied for matroids in general, as in~\cite{tamo13,westerback15,izs16}.

\subsection{Matroid fundamentals} 

Matroids have many equivalent definitions in the literature. Here, we choose to present matroids via their rank functions. Much of the contents in this section can be found in more detail in \cite{freij-hollanti17}.

\begin{definition}
\label{def:matroid_rank}
A \emph{(finite) matroid} $M=(E, \rho)$ is a finite set $E$ together with a \emph{rank function} $\rho:2^E \rightarrow \mathbb{Z}$ such that for all subsets $X,Y \subseteq E$
$$
\begin{array}{rl} 
(R.1) & 0 \leq \rho(X) \leq |X|,\\
(R.2) & X \subseteq Y \quad \Rightarrow \quad \rho(X) \leq \rho(Y),\\
(R.3) & \rho(X) + \rho(Y) \geq \rho(X \cup Y) + \rho(X \cap Y). 
\end{array}
$$ 
\end{definition}

A subset $X \subseteq E$ is called \emph{independent} if $\rho(X) = |X|$. If $X$ is independent and $\rho(X) = \rho(E)$, then $X$ is called a \emph{basis}. A subset that is not independent is called \emph{dependent}. A circuit is a minimal dependent subset of $E$, that is, a dependent set whose proper subsets are all independent. Strongly related to the rank function is the \emph{nullity function} $\eta:2^E \rightarrow \mathbb{Z}$, defined by $\eta(X) = |X| - \rho(X)$ for $X \subseteq E$. 

Any matrix $G$ over a field $\F$ generates a matroid $M_G=(E,\rho)$, where $E$ is the set of columns of $G$, and $\rho(X)$ is the rank of $G(X)$ over $\F$, where $G(X)$ denotes the submatrix of $G$ formed by the columns indexed by $X$. As elementary row operations preserve the row space of $G(X)$ for all $X\subseteq E$, it follows that row-equivalent matrices generate the same matroid.  

Thus, there is a straightforward connection between linear codes and matroids. Let $C$ be a linear code over a field $\F$. 
Then any two different generator matrices of $C$ will have the same row space by definition, so they will generate the same matroid. Therefore, without any inconsistency, we can denote the matroid associated to these generator matrices by $M_C = (E, \rho_C)$. The rank function $\rho_C$ can be defined directly from the code without referring to a generator matrix, via $\rho_C(X) = \mathrm{dim}(C|X)$ for $X \subseteq E$.

One way of defining a new matroid from an existing one is obtained by restricting the matroid to one of its subsets. For a given set $X \subseteq E$, we define the \emph{restriction} of $M$ to $X$ to be the matroid $M|X = (X, \rho_{|X})$ by $\rho_{|X} (Y) = \rho(Y)$ for all subsets $Y \subseteq X$.

Two matroids $M_1 = (E_1, \rho_1)$ and $M_2 = (E_2, \rho_2)$ are \emph{isomorphic} if there exists a bijection $\psi: E_1 \rightarrow E_2$ such that $\rho_2(\psi(X)) = \rho_1(X)$ for all subsets $X \subseteq E_1$.

\begin{definition} A matroid that is isomorphic to $M_G$ for some matrix $G$ over $\F$ is said to be \emph{representable} over $\F$. We also say that such a matroid is $\F$-representable. 
A \emph{binary} matroid is a matroid that is $\F_2$-representable.
\end{definition}

By the Critical Theorem \cite{crapo70}, the matroid $M_C$ determines the supports of a linear code $C$. Consequently, since binary codes are determined uniquely by the support of the codewords, binary matroids are in one-to-one correspondence with binary codes. This is in sharp contrast to linear codes over larger fields, where many interesting properties are not determined by the associated matroid. 
An important example of such a property is the covering radius \cite{britz05}. 

\begin{definition}
A matroid is called \emph{simple} if if has no circuits consisting of $1$ or $2$ elements. A element $e \in E$ is called a \emph{co-loop} if $\rho(E-e) < \rho(E)$. 
\end{definition}

\subsection{Fundamentals on cyclic flats}

The main tool from matroid theory in this paper are the cyclic flats. We will define them using the closure and cyclic operators.   

Let $M = (E,\rho)$ be a matroid. The \emph{closure} operator $\mathrm{cl}:2^E \rightarrow 2^E$ and \emph{cyclic} operator $\mathrm{cyc}: 2^E \rightarrow 2^E$ are defined by
$$
\begin{array}{cl}
(i) & \mathrm{cl}(X) = X \cup \{e \in E -X : \rho(X \cup e) = \rho(X)\},\\
(ii) & \mathrm{cyc}(X) = \{e \in X : \rho(X - e) = \rho(X)\}.
\end{array}
$$
A subset $X \subseteq E$ is a \emph{flat} if $\mathrm{cl}(X) = X$ and a \emph{cyclic set} if $\mathrm{cyc}(X) = X$. Therefore, $X$ is a \emph{cyclic flat} if 
$$
\rho(X \cup y) > \rho(X) \quad \hbox{and} \quad \rho(X - x) = \rho(X)
$$
for all $y \in E-X$ and $x \in X$. The collection of flats, cyclic sets, and cyclic flats of $M$ are denoted by $\mathcal{F}(M)$, $\mathcal{U}(M)$, and $\mathcal{Z}(M)$, respectively.

Some more fundamental properties of flats, cyclic sets, and cyclic flats are given in \cite{bonin08}.

The cyclic flats of a linear code $C$ of $\F^E$ can be described as sets $X \subseteq E$ such that 
$$
|C|(X \cup y)| >  |C|X| \quad \hbox{and} \quad |C|(X - x)| = |C|X|
$$ 
for all $y \in E -X$ and $x \in X$.

Before going deeper in the study of $\ZZ(M)$, we need a minimum background on poset and lattice theory. We will use the standard notation of $\wedge$ and $\vee$ for the meet and join operator, we will denote by $0_{\LL}$ and $1_{\LL}$ the bottom and top element of a lattice $\LL$, and we will denote by $\lessdot$ the covering relation, \emph{i.e.}, for $X,Y \in (\LL, \leq)$, we say that $X \lessdot Y$ if $X < Y$ and there is no $Z \in \LL$ with $X < Z < Y$.

The atoms and coatoms of a lattice $(\LL, \subseteq)$ are defined as
\begin{align*}
 A_{\LL} & = \{ X \in \LL \setminus 0_{\LL} : \nexists Y \in \LL \st 0_{\LL} \subsetneq Y \subsetneq X \}, \\
coA_{\LL} & = \{ X \in \LL \setminus 1_{\LL} : \nexists Y \in \LL \st X \subsetneq Y \subsetneq 1_{\LL} \},
\end{align*}
respectively. 

A lattice $\LL$ is said to be \emph{atomic} if every element of $\LL$ is the join of atoms. For further information about posets and lattices, we refer the reader to \cite{stanley2011}.

We can now give a crucial property of the set of cyclic flats. 

\begin{proposition}[See \cite{bonin08}]
Let $M=(E, \rho)$ be a matroid. Then
\begin{enumerate}
\item $(\ZZ(M), \subseteq)$ is a lattice with $X \vee Y = \cl (X \cup Y)$ and $X \wedge Y = \cyc(X \cap Y)$.
\item $1_{\ZZ} = \cyc(E)$ and $0_{\ZZ} = \cl(\emptyset)$.
\end{enumerate}
\end{proposition}

\subsection{Relation between LRCs and the lattice of cyclic flats}

Recently, some work has been done to emphasise the relation between cyclic flats and linear codes over finite fields. In \cite{westerback15}, the authors proved that the minimum distance can be expressed in terms of the nullity of certain cyclic flats: 

\begin{proposition}
\label{prop: d_via_cyclicflats}
Let $C$ be a non-degenerate $(n,k,d)$-code and $M=(E,\rho)$ the matroid associated to $C$. Then,
\[
d = \eta(E) +1 - \max \{ \eta(Z) : Z \in \ZZ(M) - \{E \} \}.
\]
\end{proposition}

Moreover, \cite{grezet17} gives us necessary conditions on the structure of the lattice of cyclic flats when the code and hence the matroid are binary. The key results from \cite{grezet17} are the following proposition and theorem that constrain the edges of the associated Hasse diagram. 

\begin{proposition}[\cite{grezet17}]
\label{prop:edge_rank_null}
Let $M=(E,\rho)$ be a binary matroid. Then, every $X,Y \in \ZZ(M)$ with $X \lessdot Y$ satisfy exactly one of the following:
\begin{itemize}
\item $\rho(Y)- \rho(X)=l>1$ and $\eta(Y)- \eta(X) =1$. We call such an edge in the Hasse diagram of $\ZZ(M)$ a \textit{rank edge} and label it $\rho=l$.
\item $\rho(Y)-\rho(X) = 1$ and $\eta(Y)- \eta(X)=l>1$. We call such an edge a \textit{nullity edge} and label it $\eta=l$.
\item $\rho(Y)-\rho(X) = 1$ and $\eta(Y)- \eta(X) =1$. We call such an edge a \textit{elementary edge}.
\end{itemize}
\end{proposition}

\begin{theorem}[Announced in \cite{grezet17}]
\label{thm:castle_cov_relation}
Let $C$ be a non-degenerate, binary linear $(n,k,d,r,\delta)$-LRC with $d>2$ and without replication. Let $M=(E,\rho)$ be the associated matroid. Then $\ZZ=\ZZ(M)$ satisfies the following:
\begin{enumerate}
\item $\emptyset$ and $E$ are cyclic flats.
\item Every covering relation $Z \lessdot E$ is a nullity edge labeled with a number $\geq d-1$.
\item If $\delta=2$, then for every $i \in E$, there is $X \in \ZZ$ with $i \in X$ such that $\rho(X)\leq r$.
\item If $\delta>2$, then for every $i \in E$, there is $X \in \ZZ$ with $i \in X$ such that
\begin{enumerate}
\item Every covering relation $Y \lessdot X$ is a nullity edge labeled with a number $\geq \delta -1$.
\item Every cyclic flat $Y$ with $Y \lessdot X$ has rank $\leq r-2$.
\end{enumerate}
\end{enumerate}
\end{theorem}
\begin{proof}
From non-degeneracy, it follows that $\emptyset$ and $E$ are cyclic flats. If $Z \lessdot E$, then by Proposition~\ref{prop: d_via_cyclicflats}, $d \leq \eta(E) +1 - \eta(Z)$. As $d>2$, this means that $1<d-1\leq \eta(E)-\eta(Z)$, so $Z \lessdot E$ is a nullity edge. This proves 1) and 2).

Now for every $i\in E$, there is $R\subseteq E$ with $i\in R$, $d_R\geq \delta$ and $|R|-\delta +1\leq r$. In particular, when $\delta>1$, $R$ must be a cyclic set, and its closure $X$ is a cyclic flat with $\rho(X)=\rho(R)$ and $d_X\geq d_R\geq \delta$. Now, (3) follows as $\rho(X)=\rho(R)\leq |R|-d_R +1\leq |R|-\delta+1\leq r$.

Finally, assume $\delta>2$ let $Y\lessdot X$. Since $d_{X} = d_{R} \geq \delta$, it follows from 2) that $Y \lessdot X$ is a nullity edge labeled with a number $\geq \delta -1$. Consider now the lattice $\ZZ(M|R)$. We have $0_{\ZZ(M|R)}=\emptyset$ and $1_{\ZZ(M|R)} = R$ since R is cyclic. As there is no replication, $\rho(R) \geq 2$ and $\eta(R) \geq \delta -1 \geq 2$ by Proposition~\ref{prop: d_via_cyclicflats}. It follows from Proposition \ref{prop:edge_rank_null} that there exists a cyclic flat $Z_{c} \in \ZZ(M|R)$ different from $\emptyset$ and $R$. Then it follows that $$\max \{ \eta(Z) : Z \in \ZZ(M|R) - \{R \} \} \geq 1,$$ $$\eta(R) = d_{R}+ \max \{ \eta(Z) : Z \in \ZZ(M|R) - \{R \} \} - 1 \geq \delta, $$ and $$\rho(R) = |R| - \eta(R) \leq  r-1.$$ Hence, $\rho(X) \leq r-1$ and all cyclic flats $Y \lessdot X$ have rank $\leq r-2$.

\end{proof}

\section{Lattice structure and repair properties}\label{sec:lattice}

The first part of this section is devoted to understanding how restricting to binary linear codes affects the structure of the lattice of cyclic flats. We will see that Proposition \ref{prop:edge_rank_null} has a strong impact on the structure of the atoms of $\ZZ$, which in return influences the whole lattice.  Indeed the main result of this section consists of proving that the lattice of cyclic flats has the property of being atomic. 

In the second part, we will discuss the meaning of these results for binary linear codes and LRCs. In particular, we will see that every non-degenerate binary linear $(n,k,d)$-code without replication is already a binary linear $(n,k,d,r', 2)$-LRC for a certain $r'$. Furthermore, for LRCs with $\delta >2$, we will see that these codes have a hierarchy in failure tolerance.

\subsection{Structural properties of the lattice of cyclic flats}
We will first begin by the relation between binary linear codes and the associated matroid. The following proposition is a reformulation of Proposition 8 in \cite{westerback15} together with the easy observation that, in a binary linear code, two symbols are dependent if and only if they are equal.

\begin{proposition}
\label{prop:non_degenerated}
Let $C$ be a binary linear $(n,k,d)$-code. Then $C$ is non-degenerate with no replication if and only if the associated matroid $M=(E,\rho)$ is simple and contains no co-loops. 
\end{proposition}

Now that we have establish the type of matroids that is revelant to our case, we can study the implications of Proposition \ref{prop:edge_rank_null} for the lattice of cyclic flats. We begin by an immediate consequence on the atoms of $\ZZ(M)$.

\begin{lemma}
\label{lemma:atom_iff_eta1}
Let $M=(E,\rho)$ be a simple binary matroid. Then,
\[
Z \text{ is an atom of } \ZZ(M) \text{ if and only if } \eta(Z) = 1.
\]
\end{lemma}

\begin{proof}
Since $M$ is simple, it guaranties that $\emptyset = 0_{\ZZ}$. Furthermore, it also means that, for all cyclic flats $Z \neq \emptyset$, we have $\rho(Z)>1$. Hence, by Proposition \ref{prop:edge_rank_null}, every atom $Z_{at}$ will have a rank edge, \emph{i.e.}, $\eta(Z_{at}) = 1$. 
\end{proof}

The next two lemmas link atoms in $\ZZ(M)$ to certain short circuits in $M$. 

\begin{lemma}
\label{lemma:circuit_atom}
Let $M=(E,\rho)$ be a simple binary matroid. Let $C$ be a circuit of $M$. Then $\cl(C)$ is an atom of $\ZZ(M)$ if and only if $\cl(C)=C$. 
\end{lemma}

\begin{proof}
By lemma \ref{lemma:atom_iff_eta1}, $\cl(C)$ is an atom of $\ZZ(M)$ if and only if $\eta(\cl(C))=1$. But since $C$ is a circuit, we have $\eta(C)=1$. Now $\eta(\cl(C)) = 1 + |\cl(C)-C|$. Hence $\cl(C)$ is an atom if and only if $\cl(C) = C$.
\end{proof}

\begin{lemma}
\label{lemma:circuit_closure}
Let $M=(E,\rho)$ be a simple binary matroid that contains no co-loops. Let $e \in E$ and $C$ be a circuit containing $e$ of minimal length. Then $\cl(C)=C$.
\end{lemma}

\begin{proof}
First, the existence of a circuit containing $e$ is guaranteed by the fact that $M$ contains no co-loops. Let $C$ be a minimum length circuit containing $e$. 

Now, consider a binary representation $\{x_{f} \}_{f \in M}$ of $M$. We can express $x_{e}$ by a linear combination of elements $\{x_{f} : f \in C\setminus \{e\} \}$. Since $C$ is a binary circuit, we will need all elements in $C \setminus \{e\}$ with coefficient equal to 1. Hence
\[
x_{e}=\sum\limits_{f \in C\setminus \{ e\} } x_{f}.
\]
Assume for a contradiction there exists $e' \in \cl(C) - C$. Then $x_{e'} = \sum\limits_{f \in D \subseteq C}x_{f}$. Since $M$ is binary and simple, we have $2 \leq |D| < |C|$. 

If $e \in D$, then we have found a circuit smaller than $C$ containing $e$ which is a contradiction to the minimality of $C$. 

If $e \notin D$, then
\[
x_{e}= \sum\limits_{f \in D}x_{f} + \sum\limits_{f' \in (C\setminus \{e \} ) \setminus D} x_{f'} = x_{e'} + \sum\limits_{f' \in (C\setminus \{e \} ) \setminus D} x_{f'}.
\]
Thus, $e$ is in the circuit $\{e \} \cup \{e' \} \cup ((C\setminus \{e \}) \setminus D)$ with cardinality $| \{e'\} \cup (C \setminus D)| < |C|$ by the fact that $|D|\geq 2$. Again, this is a contradiction to the minimality of $C$. Hence $\cl(C)=C$.
\end{proof}

By combining Lemma \ref{lemma:circuit_atom} and \ref{lemma:circuit_closure}, we obtain the following result. 

\begin{lemma}
\label{lemma:element_in_atom}
Let $M=(E,\rho)$ be a simple binary matroid that contains no co-loops. Then every element $e \in E$ is contained in an atom. 
\end{lemma}

\begin{theorem}
\label{thm:lcf_atomic}
Let $M=(E,\rho)$ be a simple binary matroid that contains no co-loops. Then the lattice of cyclic flats $\ZZ(M)$ is atomic.
\end{theorem}

\begin{proof}
By Lemma \ref{lemma:element_in_atom}, for every $e \in E$, there exist an atom $Z_{at}^{e} \in \ZZ(M)$ with $e \in Z_{at}^{e}$. Thus, 
\[
\bigvee\limits_{e \in M} Z_{at}^{e} \supseteq \bigcup\limits_{e \in M}Z_{at}^{e} = E.
\]
For $Y \in \ZZ(M)$, we can restrict the matroid to $M|Y=(Y,\rho)$. Since $\ZZ(M|Y) = \{ Z : Z \subseteq Y, Z \in \ZZ(M) \}$ and $M|Y$ contains no co-loops, we are back to the previous case. Hence
\[
Y = \bigvee\limits_{e \in Y} Z_{at}^{e}
\]
and this proves that $\ZZ(M)$ is atomic. 
\end{proof}
Indeed, we have proven a slightly stronger property than atomicity. Namely, any element in $\ZZ(M)$ is equal not only to the join, but also to the union of all the atoms that it contains.

\subsection{Hierarchy of failure tolerance}

By Proposition \ref{prop:non_degenerated}, non-degenerate binary linear $(n,k,d)$-codes with no replication are associated with simple binary matroids that contain no co-loops. Thus, we can reinterpret these results for this type of binary codes. Probably the most significant one is Lemma \ref{lemma:element_in_atom} stating that every element of a given matroid $M$ is contained in some atoms, meaning that every symbol of the associated code is automatically contained in a small dependent set. Furthermore, by Proposition \ref{prop: d_via_cyclicflats}, the minimum distance of an atom is 2. Thus, every symbol is directly contained in a small repair set with $\delta=2$, \emph{i.e.}, in a repair set that can correct exactly one erasure. Hence we obtain the following theorem. 

\begin{theorem}
\label{thm:delta2}
For every non-degenerate binary linear $(n,k,d)$-code $C$ with no replication, $C$ is also an $(n,k,d,r',2)$-LRC for some $r' \in \{2, \ldots , k \}$. 
\end{theorem}

\begin{proof}
Let $M=(E,\rho)$ be the associated matroid to $C$. By Proposition \ref{prop:non_degenerated}, $M$ is simple and contains no co-loops. By Lemma \ref{lemma:element_in_atom}, for all $e \in E$, there exists an atom $Z_{at}^{e} \in \ZZ(M)$ with $e \in Z_{at}^{e}$. Define $r'=\max\limits_{e \in E} \rho(Z_{at}^{e} )$. Then the collection $\{Z_{at}^{e} \}$ gives a locality $(r',\delta=2)$ to $C$. Indeed, by Proposition \ref{prop:edge_rank_null} and \ref{prop: d_via_cyclicflats} we have $\eta(Z_{at}^{e}) = 1$ and $d_{Z_{at}^{e}} = 2$, which gives us $\delta =2$. For the size criterion, we have
\begin{align*}
|Z_{at}^{e}| &= \rho(Z_{at}^{e}) + \eta(Z_{at}^{e}) = \rho(Z_{at}^{e}) + \delta -1 \\
& \leq \max\limits_{e\in E} \{ \rho(Z_{at}^{e}) \} + \delta -1 = r' + \delta -1.
\end{align*}
\end{proof}

Now, if we want to be able to correct more than one erasure, then the repair sets cannot be atoms of $\ZZ(M)$ as these have $d_{Z}=2$. They have to be at least one level above some atoms. However, the previous theorem still holds, meaning that for every symbol, there is also an atom containing it. Thus, we get a natural hierarchy in failure tolerance. If one node fails, then we can contact the close-by nodes in the atom to repair it. If more nodes fail, but no more than $\delta-1$, we can contact other repair sets to fix them. And if more than $\delta -1$ nodes fail, then we need to use the global properties of the code. 

Moreover, by the remark following Theorem \ref{thm:lcf_atomic}, it follows that repair sets are unions of all the atoms below them. Since the collection of repair sets contains every symbol, we can choose the collection of atoms that will give us the $(r',2)$ locality to be inside repair sets. The following corollary summaries the previous observations in one statement. 

\begin{corollary}
Let $C$ be a non-degenerate binary linear $(n,k,d,r,\delta)$-LRC with no replication and with $\delta >2$. Let $\{R_{i} \}_{i \in I}$ be the list of repair sets. Then, there exists a collection of sets $\{ X_{j} \}_{j \in J}$ such that for all $X_{j}$, there exists $R_{i}$ with $X_{j} \subsetneq R_{i}$ such that $C$ is also an $(n,k,d,r',2)$-LRC. 
\end{corollary}

From a practical point of view, this reinforces the usefulness of the failure tolerance hierarchy. For example, suppose that the symbol $e \in E$ fails. We can start by downloading nodes in the atom $Z_{at}^{e}$ in order to repair $e$. Now, if we realize that some other nodes in $Z_{at}^{e}$ have failed as well, we can keep the part that we already downloaded from $Z_{at}^{e}$ and contact more nodes in the corresponding repair set to repair all the failed symbols in $Z_{at}^{e}$. Thus, we do not have to restart from the beginning if we find out during the repair process that a small amount of other nodes have failed as well.

\section{Improving the singleton-type bound for $\delta>2$}

The goal of this section is to improve the existing bound for non-degenerate linear $(n,k,d,r,\delta)$-codes $C$ when the codes are binary, contain no replication and $\delta >2$. It has been proven in \cite{kamath14} that, for a linear $(n,k,d,r,\delta)$-code over $\F_{q}$, we have
\begin{equation}
d \leq n-k+1-\left( \left\lceil \frac{k}{r} \right\rceil -1 \right) (\delta -1) \label{eq:old_b}
\end{equation}

Since the predominant parameter of this study will be the rank of the repair sets instead of the size, we assume from now on that repair sets are maximal after fixing its rank or, in matroid terms, that repair sets are cyclic flats. We will denote these repair sets by $Z_{i}$ instead of $R_{i}$ to avoid confusion.

Remember that Proposition \ref{prop: d_via_cyclicflats} links the minimum distance of a code to the coatoms among the cyclic flats. So we would like to construct a cyclic flat that is as close as possible to the coatoms level to give an accurate lower bound on $$\max \{ \eta(Z) : Z \in \ZZ(M) - \{E \} \}.$$ We will do this by creating a chain in $\ZZ(M)$ made of joins of repair sets and we will call these type of chains \textit{repair-sets}-chain or, for short, \textit{rps}-chain.

\begin{definition}
Let $C$ be a non-degenerate $(n,k,d,r,\delta)$-code and $\{ Z_{i} \}_{i \in I}$ the collection of repair sets. Let $M=(E,\rho)$ the associated matroid. An \emph{\textit{rps}-chain} 
\[
\emptyset = Y_{0} \subsetneq Y_{1} \subsetneq \ldots  \subsetneq Y_{m}=E
\]
is a chain in $\ZZ(M)$ defined inductively by
\begin{enumerate}
\item Let $Y_{0}= \emptyset$.
\item Given $Y_{i-1} \subsetneq E$, we choose $x_{i} \in E \setminus Y_{i-1}$ and $Z_{i}$ with $x_{i} \in Z_{i}$ arbitrarily, and assign $Y_{i} = Y_{i-1} \vee Z_{i}$.
\item If $Y_{i} = E$, we set $m = i$.
\end{enumerate}
\end{definition}

Notice that this chain is not uniquely defined since we can choose symbols and corresponding repair sets freely. 

Now that we have created this chain, we are interested in how the rank and the nullity can increase at each step. This will be useful when evaluating the rank and the nullity at the end of the chain. First, we will define a new parameter that represents the maximum rank of a repair set. 

\begin{definition}
Let $C$ be an $(n,k,d,r,\delta)$-LRC. Let $\{ Z_{i} \}_{i \in I}$ be the list of repair sets and $M=(E,\rho)$ the associated matroid to $C$. We define $\ell$ to be 
\[
\ell := \max\limits_{i \in I} \rho(Z_{i}) 
\]
\end{definition}

\begin{lemma}
\label{lemma:rpschain_stepbound}
Let $C$ be a non-degenerate, binary linear $(n,k,d,r,\delta)$-LRC with $\delta >2$ and without replication and $M=(E,\rho)$ the associated matroid. Then, every \textit{rps}-chain $(Y_{i})_{i=0}^{m}$ has the following properties.
\begin{enumerate}
\item $\rho(Y_{i}) - \rho(Y_{i-1}) \leq \ell$ for all $i=1, \ldots , m $.
\item $\eta(Y_{i}) - \eta(Y_{i-1}) \geq \delta -1$ for all $i=2, \ldots , m$.
\end{enumerate}
\end{lemma}

\begin{proof}
Let $\{Z_{i} \}_{i \in I}$ be the collection of repair sets and $(Y_{i})_{i=0}^{m}$ an arbitrary \textit{rps}-chain. By rank axioms, closure property and by Proposition \ref{prop:bounds_repairsets}, we have
\[
\rho(Y_{i})  \leq \rho(Y_{i-1}) + \rho(Z_{i}) - \rho(Y_{i-1} \cap Z_{i}) \leq \rho(Y_{i-1}) + \ell
\]
for all $i=1, \ldots , m$.

Moreover, by nullity properties coming from rank axioms and by Proposition \ref{prop:bounds_repairsets}, we have
\begin{align*}
\eta(Y_{i}) & \geq \eta(Y_{i-1} \cup Z_{i}) \geq \eta(Y_{i-1}) + \eta(Z_{i}) - \eta(Y_{i-1} \cap Z_{i}) \\
& \geq \eta(Y_{i-1}) + \eta(Z_{i}) - \max \{ \eta(Z) : Z \in \ZZ(Z_{i}) - Z_{i} \} \\
& \geq \eta(Y_{i-1}) + d_{Z_{i}} -1 \geq \eta(Y_{i-1}) + \delta -1 \\
\end{align*}
for all $i=2, \ldots , m$.
\end{proof}

We are now ready to present the first of our main results, a Singleton-type bound on the parameters $n, k, d, \ell$ and $\delta$. This will later be improved in Section \ref{sec:alpha}.

\begin{theorem}
\label{thm:new_singleton_bound}
Let $C$ be a non-degenerate, binary linear $(n,k,d,r,\delta)$-LRC with $\delta >2$ and without replication. Then,
\begin{equation}
d \leq n- k - \left( \left\lceil \frac{k}{ \ell } \right\rceil -1 \right) (\delta -1).\label{eq:f_b}
\end{equation}
\end{theorem}

\begin{proof}
Let $M=M(E, \rho)$ be the matroid associated to $C$, $\{ Z_{i} \}_{i \in I}$ the collection of repair sets and $(Y_{i})_{i=0}^{m}$ an \textit{rps}-chain. Our goal is to obtain a lower bound on $\max \{ \eta(Z) : Z \in \ZZ(M) - \{ E \} \}$ since it will give us a upper bound for the minimum distance by Proposition \ref{prop: d_via_cyclicflats}. 

To do this, we will give a lower bound on the nullity of the penultimate set in the \textit{rps}-chain using Lemma \ref{lemma:rpschain_stepbound}. Indeed, we have
\[
\eta(Y_{m-1}) \geq (m-2) (\delta -1) + \eta(Z_{1}).
\]

Now, by Theorem \ref{thm:lcf_atomic}, there exists an atom $Z_{at}$ below $Z_{i}$. Then, $\max \{ \eta(Z) : Z \in \ZZ(Z_{i}) - Z_{i} \} \geq \eta(Z_{at})=1$. Thus, by Proposition \ref{prop: d_via_cyclicflats}, we have
\begin{equation}
\eta(Z_{i}) \geq \delta -1 + \max \{ \eta(Z) : Z \in \ZZ(Z_{i}) - Z_{i} \} \geq \delta. \label{eq:1}
\end{equation}

Hence,
\[
\eta(Y_{m-1}) \geq (m-2) (\delta -1) + \delta = (m-1) (\delta -1) +1.
\]

To evaluate $m$, we use the upper bound on the rank. We have
\[
k=\rho(Y_{m}) \leq m \ell  \Rightarrow m \geq \left\lceil \frac{k}{ \ell } \right\rceil.
\]

Combining the two previous results, we get
\[
\eta(Y_{m-1}) \geq \left( \left\lceil \frac{k}{ \ell } \right\rceil -1 \right) (\delta -1) +1.
\]

Applying this and Proposition \ref{prop: d_via_cyclicflats}, we obtain
\begin{align*}
d_{C} & = n-k+1 - \max \{ \eta(Z) : Z \in \ZZ(M)-\{E\} \} \\
& \leq n-k +1 - \eta(Y_{m-1}) \\
& \leq n-k - \left( \left\lceil \frac{k}{ \ell } \right\rceil -1 \right) (\delta -1).
\end{align*}

\end{proof}

Even if this bound looks graphically the same as the previous known bound with an extra $-1$, the meaning is truly different since the ceiling $\left\lceil \frac{k}{ \ell } \right\rceil$ depends on the maximum rank $\ell$ of the repair sets and not on the parameter $r$ related to the size. To illustrate this, and to make the comparison easier, we will give a bound on the rank with $r$. This will also allow us to give as a corollary a new version of the last bound that only depends on $n, k, d, r$ and $\delta$.

\begin{proposition}
\label{prop:bounds_repairsets}
Let $C$ be a non-degenerate, binary linear $(n,k,d,r,\delta)$-LRC with $\delta >2$ and without replication. Let $\{Z_{i} \}_{i \in I}$ be the collection of repair sets. Then,
\[
\rho(Z_{i}) \leq r-1
\]
for all $i \in I$. In particular, this implies that $\ell \leq r-1$. 
\end{proposition}

\begin{proof}
We already saw in Equation \eqref{eq:1} that $\eta(Z_{i}) \geq \delta$. Now, using the definition of a repair set and the lower bound for $\eta(Z_{i})$, we get
\[
\rho(Z_{i}) = |Z_{i}| - \eta(Z_{i}) \leq r-1.
\]
\end{proof}

We can now reformulate the bound in Theorem \ref{thm:new_singleton_bound}:

\begin{corollary}
Let $C$ be a non-degenerate, binary linear $(n,k,d,r,\delta)$-LRC with $\delta >2$ and without replication. Then,
\[
d \leq n- k - \left( \left\lceil \frac{k}{ r-1 } \right\rceil -1 \right) (\delta -1)
\]
\end{corollary}

Compared to the previous bound \eqref{eq:old_b}, we have improved by dropping the $+1$ and by replacing $r$ by $r-1$.

\section{Tool for achieving a better bound}
\label{sec:alpha}

This section is an attempt to be more precise on the evaluation of \textit{rps}-chains which will lead to a further improvement in the bound. We will first discuss about the proof of Theorem~\ref{thm:new_singleton_bound} and Lemma~\ref{lemma:rpschain_stepbound}. 

The main ingredient in the previous section is the evaluation of the rank and nullity difference at each step on \textit{rps}-chains in Lemma \ref{lemma:rpschain_stepbound}. Indeed, if $\rho(Y_{i}) - \rho(Y_{i-1}) = \ell$, then we must have $Y_{i-1} \cap Z_{i} = \emptyset$ and $\rho(Z_{i})=\ell$. In particular, the intersection condition means that $\eta(Y_{i}) - \eta(Y_{i-1}) \neq \delta -1$.  Hence, there is no code nor an \textit{rps}-chain that can simultaneously achieve both bounds in Lemma \ref{lemma:rpschain_stepbound}.

The idea is to introduce an indicator function that will capture when the intersection is a coatom of a repair set. To be more concise, we will denote by $\Aa_{i}$ the event when $Y_{i-1}\cap Z_{i} \in coA_{\ZZ(M|Z_{i})}$. 

First, this is a necessary condition to have $\eta(Y_{i}) - \eta(Y_{i-1}) = \delta-1$. Secondly, this will also imply that $\rho(Y_{i}) - \rho(Y_{i-1}) = 1$ since every covering relation of a repair set has a nullity edge. Thus, this indicator will be able to separate the two previous extremal cases impossible to reach at the same time and allows us to have a more precise view of each of the chain step. Thus, the following lemma with the new indicator is an improvement of Lemma \ref{lemma:rpschain_stepbound}.

\begin{lemma}
\label{lemma:rpschain_indicator_stepbound}
Let $C$ be a non-degenerate, binary linear $(n,k,d,r,\delta)$-LRC with $\delta >2$ and without replication, and let $M=(E,\rho)$ be the associated matroid. Let $\{Z_{i}\}_{i\in I}$ be the collection of repair sets and $(Y_{i})_{i=0}^{m}$ an associated \textit{rps}-chain. Then $(Y_{i})_{i=0}^{m}$ has the following properties:
\begin{enumerate}
\item $\rho(Y_{i}) - \rho(Y_{i-1}) \leq \ell -( \ell -1)\mathbbm{1}_{\Aa_{i} }$ for all $i=2, \ldots , m $.
\item $\eta(Y_{i}) - \eta(Y_{i-1}) \geq \delta -\mathbbm{1}_{\Aa_{i} }$ for all $i=2, \ldots , m$.
\end{enumerate}
\end{lemma}

\begin{proof}
Let $i\in \{ 2, \ldots , m\}$. We will begin by proving the upper bound for the rank. By rank axioms, closure property and by Proposition \ref{prop:bounds_repairsets}, we have
\[
\rho(Y_{i}) - \rho(Y_{i-1}) \leq \rho(Z_{i}) - \rho(Y_{i-1} \cap Z_{i}) \leq \ell - \rho(Y_{i-1} \cap Z_{i}).
\]
When the intersection is not an coatom, we obtain the same bound as in Lemma \ref{lemma:rpschain_stepbound}. However, if $Y_{i-1}\cap Z_{i} \in coA_{\ZZ(M|Z_{i})} $, then, by Theorem \ref{thm:castle_cov_relation}.4 , we have $\rho(Y_{i-1}\cap Z_{i}) = \rho(Z_{i})-1$. Thus $\rho(Y_{i}) - \rho(Y_{i-1}) =1$. Hence, we have
\[
\rho(Y_{i}) - \rho(Y_{i-1}) \leq \ell -( \ell -1)\mathbbm{1}_{\Aa_{i} }.
\]

We will now prove the lower bound for the nullity. By nullity properties coming from rank axioms, we have
\[
\eta(Y_{i}) - \eta(Y_{i-1}) \geq \eta(Z_{i}) - \eta(Y_{i-1} \cap Z_{i}).
\]
If $Y_{i-1}\cap Z_{i} \in coA_{\ZZ(M|Z_{i})}$, then $\eta(Z_{i}) - \eta(Y_{i-1}\cap Z_{i}) \geq d_{Z_{i}} - 1 \geq \delta -1$. Thus, we obtain the same bound as in Lemma \ref{lemma:rpschain_stepbound}. However, if $Y_{i-1}\cap Z_{i} \notin coA_{\ZZ(M|Z_{i})}$, then it implies that 
\begin{align*}
\eta(Y_{i-1}\cap Z_{i}) & \leq \max \{ \eta(Z) : Z \in coA_{\ZZ(M|Z_{i})} \} -1 \\
	& = \max \{ \eta(Z) : Z \in \ZZ(M|Z_{i}) \} -1.
\end{align*}
Therefore, 
\[
\eta(Z_{i}) -\eta(Y_{i-1}\cap Z_{i}) \geq d_{Z_{i}} \geq \delta,
\]
and
\[
\eta(Y_{i}) - \eta(Y_{i-1}) \geq \delta -\mathbbm{1}_{\Aa_{i} }.
\]
\end{proof}

In order to use Lemma \ref{lemma:rpschain_indicator_stepbound} to get a Singleton-type bound with these assumptions, we need to define a new parameter that will count the number of times the intersection $Y_{i-1} \cap Z_{i}$ is a coatom of $Z_{i}$. Since we also want to compare different codes, we will focus on the proportion of atom intersections instead of the actual number of atom intersections. 

\begin{definition}
Let $(Y_{i})_{i=0}^{m}$ be an \textit{rps}-chain. Define $0 \leq \alpha \leq 1$ by $\alpha m = \# \{ i : Y_{i-1} \cap Z_{i} \in coA_{\ZZ(M|Z_{i})} \}$. We say that $(Y_{i})_{i=0}^{m}$ is an \emph{\textit{rps$_{\alpha}$-}chain}. 
\end{definition}

We can now derive a new Singleton-type bound with the extra parameter $\alpha$. 

\begin{theorem}\label{thm:alpha}
\label{thm:new_singleton_bound_indicator}
Let $C$ be a non-degenerate, binary linear $(n,k,d,r,\delta)$-LRC with $\delta >2$ and without replication. Let $\alpha\in[0,1]$ be such that $C$ has an \textit{rps$_{\alpha}$-}chain. Then,
\[
d \leq n- k +1 +\delta - \left\lceil \left\lceil \frac{k}{ \ell- (\ell -1)\alpha} \right\rceil (\delta - \alpha ) \right\rceil
\]
\end{theorem}

\begin{proof}
Let $M=M(E, \rho)$ be the matroid associated to $C$, $\{ Z_{i} \}_{i \in I}$ the collection of repair sets and $(Y_{i})_{i=0}^{m}$ an \textit{rps$_{\alpha}$-}chain. We can use Lemma \ref{lemma:rpschain_indicator_stepbound} to get a lower bound on $m$ and $\eta(Y_{m-1})$. Now, the indicator $\mathbbm{1}_{\Aa_{i}}=1$ exactly $\alpha m$ times in the \textit{rps$_{\alpha}$-}chain. We first need to get a lower bound on $m$. We have,
\[
k=\rho(Y_{m}) \leq  \ell (m-\alpha m) + \alpha m = m( \ell - (\ell -1) \alpha )
\]
Thus,
\[
m \geq \left\lceil \frac{k}{\ell - (\ell -1) \alpha} \right\rceil
\]
Next, we can obtain a lower bound on $\eta(Y_{m-1})$. We have,
\begin{align*}
\eta(Y_{m-1}) & \geq \delta (m-1) - \alpha m = m(\delta - \alpha) - \delta \\
& \geq  \left\lceil \left\lceil \frac{k}{\ell - (\ell -1) \alpha } \right\rceil (\delta - \alpha ) \right\rceil - \delta
\end{align*}
Finally, by Proposition \ref{prop: d_via_cyclicflats}, we obtain
\[
d \leq n- k +1 +\delta - \left\lceil \left\lceil \frac{k}{\ell - (\ell -1) \alpha} \right\rceil (\delta - \alpha ) \right\rceil
\]
\end{proof}

Since the bound is valid for all $\alpha$, we can optimize $\alpha$ to get a bound for all type of \textit{rps}-chain. 

\begin{theorem}
\label{thm:new_dsb_noalpha}
Let $C$ be a non-degenerate, binary linear $(n,k,d,r,\delta)$-LRC with $\delta >2$ and without replication. Then, 
\[
d \leq n-k+1- \left( \left\lceil\frac{k}{ \ell }\right\rceil -1\right) \delta +\mathbbm{1}_{ \ell | (k-1)}.
\]
\end{theorem}
\begin{proof}
Select any \textit{rps}-chain in $C$, and observe that this is an \textit{rps$_{\bar{\alpha}}$-}chain for some $\bar{\alpha}\in[0,1]$. By Theorem~\ref{thm:alpha}, we then have
\begin{align*}
d & \leq n- k +1 +\delta - \left\lceil \left\lceil \frac{k}{ \ell -( \ell -1)\bar{\alpha}} \right\rceil (\delta - \bar{\alpha} ) \right\rceil\\
& \leq n- k +1 +\delta - \min_{\alpha\in[0,1]}\left\lceil \left\lceil \frac{k}{ \ell - (\ell -1)\alpha} \right\rceil (\delta - \alpha ) \right\rceil.
\end{align*}
For fixed value of $h_\alpha\defeq \left\lceil \frac{k}{\ell- (\ell -1)\alpha} \right\rceil$, the expression $h_\alpha(\delta-\alpha)$ is clearly minimized when $\alpha$ takes its largest possible value, \ie when $\frac{k}{\ell- (\ell -1)\alpha}=h_\alpha$, so $\alpha=\frac{\ell }{\ell -1}-\frac{k}{h(\ell -1)}$. Now, when $0\leq\alpha\leq 1$, $h_\alpha$ can take any integer value between $\frac{k}{\ell}$ and $k$. So we can rewrite the bound as 
\begin{align*}d&\leq n-k+1+\delta-\min_{h\in\Z\cap[\frac{k}{\ell}, k]}\left\lceil h\left(\delta-\frac{\ell}{\ell -1}+\frac{k}{h(\ell -1)}\right)\right\rceil\\
&= n-k+1+\delta-\min_{h\in\Z\cap[\frac{k}{\ell}, k]}\left\lceil h\left(\delta-\frac{\ell}{\ell -1}\right)+\frac{k}{\ell -1}\right\rceil.
\end{align*}
The coefficient of $h$ is clearly positive, so the minimum is achieved when $h=\left\lceil\frac{k}{\ell }\right\rceil$. Hence, we get the bound 
\begin{align*}
d & \leq n-k+1+\delta-\left\lceil \left\lceil\frac{k}{\ell }\right\rceil\left(\delta-\frac{\ell}{\ell -1}\right) +\frac{k}{\ell -1}\right\rceil \\
& = n-k+1+\delta- \left\lceil\frac{k}{\ell }\right\rceil\delta \\ & \,\,\, - \left\lceil \frac{1}{\ell -1}\left(k - \ell \left\lceil\frac{k}{\ell }\right\rceil\right) \right\rceil.\end{align*}
It is straightforward to verify that the term $\left\lceil \frac{1}{\ell -1}\left(k - \ell \left\lceil\frac{k}{\ell }\right\rceil\right) \right\rceil$ evaluates to $-1$ if $k=1\mod\ell$, and $0$ otherwise. Thus, we can write the bound as
\[d\leq n-k+1- \left( \left\lceil\frac{k}{\ell }\right\rceil -1 \right)\delta +\mathbbm{1}_{\ell |(k-1)}.\]
\end{proof}

Now, since we have two bounds with the same assumptions, we can compare them and find the minimum of the two, to get the best bound. It turns out that the latter of the two bounds is always at least as strong as the first one, expect when $r=k$. In that case only, the first bound is stronger, and the difference between the two bounds is one. Hence, we obtain the following global bound. 

\begin{theorem}
\label{thm:bound_ldelta}
Let $C$ be a non-degenerate, binary linear $(n,k,d,r,\delta)$-LRC with $\delta >2$ and without replication. Then, 
\begin{equation}
d \leq n-k+1- \left( \left\lceil\frac{k}{\ell }\right\rceil -1 \right) \delta +\mathbbm{1}_{\ell |(k-1) \; \mathrm{and} \; (\ell +1) \neq k}.
\label{eq:new_l_b}
\end{equation}

\end{theorem}

We can apply the bound from Proposition \ref{prop:bounds_repairsets} to obtain a bound on the parameters $n,k,d,r$ and $\delta$ only. 

\begin{corollary}
\label{cor:best_bound}
Let $C$ be a non-degenerate, binary linear $(n,k,d,r,\delta)$-LRC with $\delta >2$ and without replication. Then, 
\[
d \leq n-k+1- \left( \left\lceil\frac{k}{r-1 }\right\rceil -1 \right) \delta +\mathbbm{1}_{(r-1) |(k-1) \; \mathrm{and} \; r \neq k}
\]
\end{corollary} 

The following graph is a comparison of the previous known bound \eqref{eq:old_b} and the new one from Corollary \ref{cor:best_bound} for two different values of $r$. We can see that the new bound is always better than (or equivalently, smaller) or equal to the previous bound.

\begin{figure}[H]
\centering
\includegraphics[height=6cm]{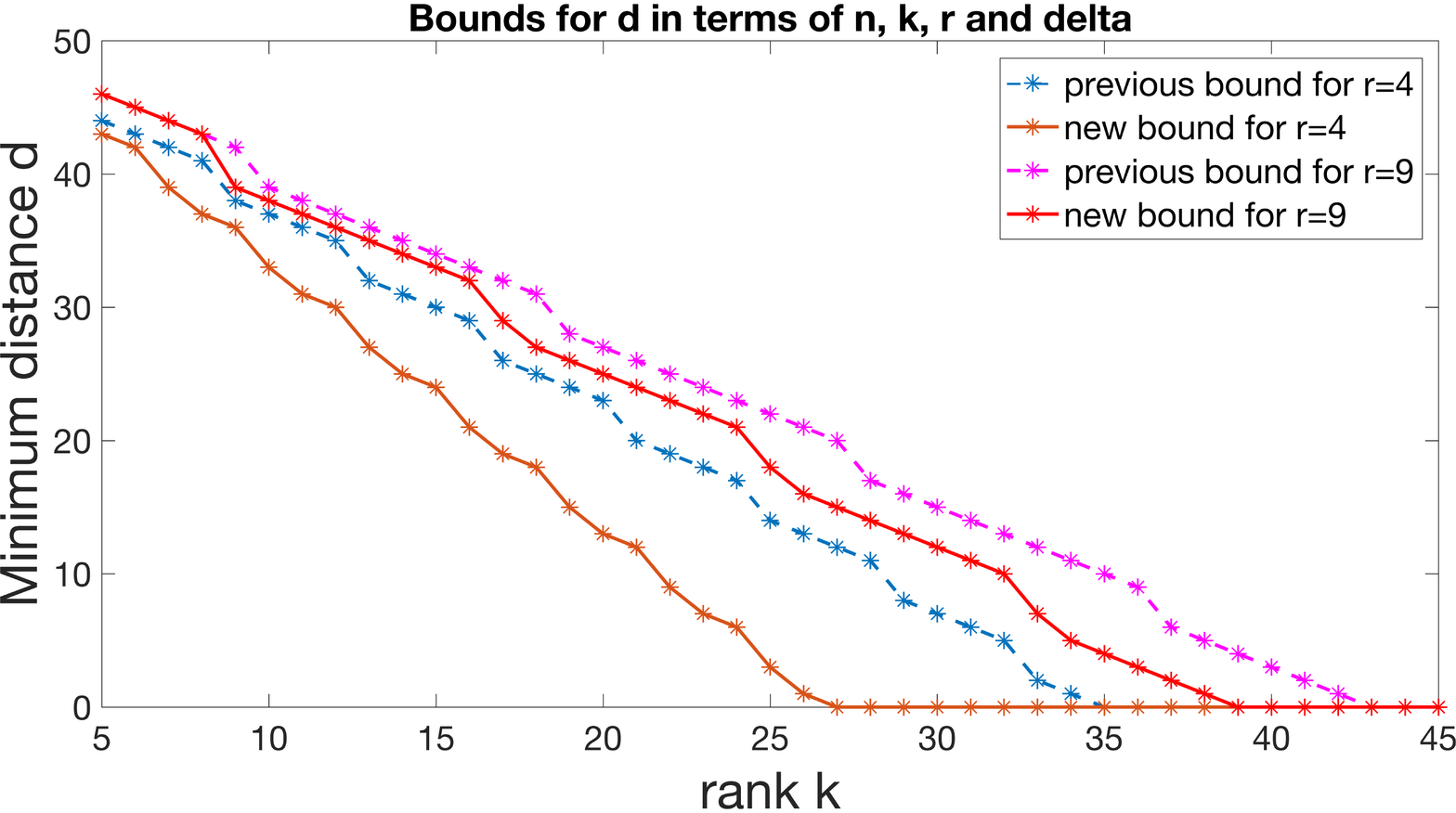}
\label{graph_comp}
\caption{Comparison of the previous bound and \\ the new bound for $n=50$ and $\delta =3$.}
\end{figure}

Finally, we provide one small example that achieves the bound from Corollary \ref{cor:best_bound}. 

\begin{example}
\label{ex1}
Let $C$ be the binary linear $(10,4,4)$-code given by the following generator matrix:
\[ G=\SmallMatrix{
1&0&0&0&1&0&1&1&1&1\\
0&1&0&0&1&1&0&1&1&1\\
0&0&1&0&0&1&0&1&0&1\\
0&0&0&1&0&0&1&0&1&1\\
}.
\]
We can define three repair sets by their corresponding columns in $G$.
\[
Z_{1}=\{1,2,3,5,6,8\}, \quad Z_{2}=\{2,3,6,7,9,10\}, \quad \nd
\]
\[
Z_{3}=\{1,4,6,7,8,10\}.
\]
For all $i \in \{1,2,3 \}$, we have
\[
|Z_{i}| = 6, \quad \rho(Z_{i})=3, \quad \nd  \quad d_{Z_{i}} = 3.
\]
Hence we obtain a binary linear $(10,4,4,4,3)$-LRC that achieves the bound from Corollary \ref{cor:best_bound}. 
\end{example}

First, even if this code has not the most interesting parameters, mainly because the rate of the code is under one-half,  it proves the achievability of the bound. Secondly, it illustrates perfectly the notion of failure tolerance hierarchy mentioned in Section \ref{sec:lattice}. To avoid confusion, we denote now by $\{c_{1}, c_{2}, \ldots , c_{10} \}$ the set of columns of $G$. Suppose that the column $c_{1}$ fails. Then, we can repair it by summing columns $c_{2}$ and $c_{4}$ in the corresponding atom. If $c_{1}$ and $c_{2}$ fail, then we form a basis in the repair set $Z_{1}$ and we repair both columns via the basis. For example the set $\{c_{3}, c_{6}, c_{8} \}$ is a basis of the first repair set $Z_{1}$, and $c_{1} = c_{6} \oplus c_{8}$ and $c_{2} = c_{3} \oplus c_{6}$. Finally, if the three columns $c_{1}, c_{2}$ and $c_{5}$ fail, we use the global correcting properties of the code. Here, we can choose the basis $\{c_{3},c_{4},c_{6},c_{7} \}$ and we have $c_{1} = c_{4} \oplus c_{7}$, $c_{2} = c_{3} \oplus c_{6}$, and $c_{5} = c_{3} \oplus c_{4} \oplus c_{6} \oplus c_{7}$.

The most commonly used field-dependent distance bound for LRCs is the Cadambe--Mazumdar bound~\cite{cadambe15}, which is only stated without reference to the local minimum distance (or equivalently, for $\delta=2$). It can straightforwardly be generalized to codes tolerating more than one local erasure. To the best of our knowledge, this generalization has not previously appeared in the literature. An unfortunate conflict of notation in the literature caused an erroneous version of Proposition 6 to occur in \cite[Proposition 6]{izs18}. In~\cite{rawat15}, a similar bound was presented when more than one erasure is tolerated, but the system assumptions used there are quite different from those in the present paper, and hence the bounds are not directly comparable,

\begin{proposition}
\label{thm:CMdelta}
Let $C$ be a non-degenerate linear $(n,k,d,r,\delta)$-LRC over $\mathbb{F}_q$. Then,
\begin{small}
\begin{equation}\label{eq:CMdelta}
k \leq \min\limits_{t \in \Z_{+}} \left[ t r + k_{opt}^{(q)}(n-t(r+\delta -1), d) \right],
\end{equation}
\end{small} where 
$k_{opt}^{(q)}(n,d)$ is the maximum rank of a linear code of length $n$ and minimum distance $d$ over $\mathbb{F}_q$. 
\end{proposition}

However, the determination of $k_{opt}^{(q)}$ is a classical open problem in coding theory. Moreover, even given a formula for $k_{opt}^{(q)}$, evaluating \eqref{eq:CMdelta}
may be a tedious task. In that sense, the bound (\ref{eq:new_l_b}) is more explicit than this one. 

In order to compare the bounds (\ref{eq:new_l_b}) and (\ref{eq:CMdelta}) for binary codes, we choose to estimate $k_{opt}$ in (\ref{eq:CMdelta}) via the Plotkin bound for binary linear codes with specific values of $n,d,r,\delta$ and $\ell$, and denote by $k_{max}$ the maximum dimension given by (\ref{eq:CMdelta}). Then we have, on one hand
 \begin{small}
\begin{eqnarray}
&&k+  \left\lceil\frac{k}{\ell }\right\rceil  \delta - \mathbbm{1}_{\ell |(k-1) \; \mathrm{and} \; l \neq k-1}\nonumber\\
&\stackrel{(\ref{eq:CMdelta}) }{\leq} &k_{max} +  \left\lceil\frac{k_{max}}{\ell }\right\rceil  \delta - \mathbbm{1}_{\ell |(k_{max}-1) \; \mathrm{and} \; l \neq k_{max}-1}\label{eq:comp_value}
\end{eqnarray}
\end{small} 
and on the other hand
 \begin{small}
\begin{eqnarray}
k + \left\lceil\frac{k}{\ell }\right\rceil  \delta - \mathbbm{1}_{\ell |(k-1) \; \mathrm{and} \; l \neq k-1}\stackrel{\textrm{Thm.\ref{thm:new_dsb_noalpha} }}{\leq}  n-d+1+\delta\,,
\label{eq:comp_value2}
\end{eqnarray}
\end{small} 
so the left hand side in \eqref{eq:comp_value} and \eqref{eq:comp_value2} is bounded by the minimum of the right hand sides of these inequalities. The best bound will be  the one that gives the minimum of the two alternative right hand sides. This allows for a partial but computable comparison of \eqref{eq:new_l_b} and \eqref{eq:CMdelta}. Using this method, we found parameters values for which \eqref{eq:comp_value} is tighter than \eqref{eq:comp_value2} and vice-versa. In particular, the parameters from example \ref{ex1} achieve \eqref{eq:comp_value2} with equality while it is one off in \eqref{eq:comp_value}. This shows that for certain parameters values, Theorem \ref{thm:bound_ldelta} is stronger than Proposition \ref{thm:CMdelta}. In \cite{izs18}, it was erroneously claimed that Proposition \ref{thm:CMdelta} is always stronger than Theorem \ref{thm:bound_ldelta}, due to an error in the statement of Proposition \ref{thm:CMdelta}.


However, this work is only a first step toward an explicit bound for binary LRCs and further improvements can be done by exploiting the techniques developed in this paper, via more detailed study of the cyclic flats. Such improvements can be made to give stronger versions both of  Theorem \ref{thm:bound_ldelta} and of Proposition \ref{thm:CMdelta}. Due to interest of space, this is left to an extended version of this article.

\section{Conclusion}
We have derived new explicit tradeoffs between the local and global minimum distances of binary locally repairable codes, together with an example of a code meeting this bound with equality. Applications to distributed data storage were discussed, with the motivation to decrease computational complexity thanks to a small field size. Moreover, we demonstrated an intrinsic hierarchy of repair sets that naturally occurs for binary linear storage codes. 

\textbf{Acknowledgment.}
This work is supported in part by the Academy of Finland (grants \#276031, \#282938 and \#303819 to C. Hollanti), and by the TU Munich -- Institute for Advanced Study, funded by the German Excellence Initiative and the EU 7th Framework Programme (grant \#291763).

\bibliographystyle{IEEEtran}
\bibliography{references}

\end{document}